\documentclass[showkeys,showpacs,preprintnumbers,amsmath,twocolumn,amssymb,prl]{revtex4}

\usepackage{graphicx}
\usepackage{dcolumn}
\usepackage{bm}

\begin{document}

\preprint{APS/123-QED}

\title{Fractional quantum Hall effect and electron correlations in partially
filled first excited Landau level}
\author{George E. Simion}
\author{John J. Quinn}
\affiliation{Department of Physics, University of Tennessee,
Knoxville, TN 37996, USA}

\begin{abstract}
We present a quantitative study of most prominent incompressible
quantum Hall states in the partially filled first excited Landau
level (LL1) which have been recently studied experimentally by Choi
et al. The pseudopotential describing the electron - electron
interaction in LL1 is harmonic at short range. It produces a series
of incompressible states which is different from its LL0
counterpart. The numerical data indicate that the most prominent
states $\nu=\frac{5}{2}$, $\frac{7}{3}$, and $\frac{8}{3}$ are not
produced by Laughlin correlated electrons, but result from a
tendency of electrons to form pairs or larger clusters which
eventually become Laughlin correlated. States with smaller gaps at
filling factors $\frac{14}{5}$, $\frac{16}{7}$, $\frac{11}{5}$,
$\frac{19}{7}$ are Laughlin correlated electron or hole states and
fit Jain's sequence of filled $\rm{CF}^4$ levels.
\end{abstract}
\date{\today}

\maketitle

The possibility of using non-Abelian quasiparticle excitations in
quantum computing has led to a revival of interest in the fractional
quantum Hall (FQH) states of the first excited Landau level (LL1)
\cite{dassarma_QC, dassarma_QC_PhysToday, freedman_QC, kitaev_QC,
bonesteel_QC}.  The FQH states of LL1 are quite different from those
of LL0, and not nearly as well understood \cite{MooreReadNuclPhys91,
GreiterWenWilczekPRL91, HaldaneRezayiPRL88,
ReadRezayiPRB99,WojsPRB01, WojsQuinnPRB05,DasSarma52}. Before making
use of the properties of non-Abelian QPs of LL1 (of a very special
model three-body interaction) for practical purposes, it seems
worthwhile to study the incompressible quantum liquid (IQL) states
and their QPs in LL1 using realistic Coulomb interactions. A recent
careful study of the temperature dependence of the minimum
longitudinal conductivity \cite{LL1_Exp} noted that the most robust
FQH states in LL1 occur at $\nu_1=\nu-2=1/2$, $1/3$, $1/5$ and $2/7$
and their electron-hole (e-h) conjugates. It was emphasized that
this was in sharp contrast to LL0, where $\nu=1/3,2/5,3/7$ and their
e-h conjugate were the most prominent incompressible states, but no
discussion was given of why this difference occurred. In this paper
we present an explanation of why correlations in LL0 and LL1 are
different and result in different sets of prominent FQH states. We
support our explanation with numerical results for energy spectra
and for the probability $P(\mathcal R)$ that the ground state
contains pairs with relative angular momentum $\mathcal
R=1,3,5\ldots$. We restrict our consideration to states that are are
fully spin polarized for the sake of simplicity, and because they
appear to yield agreement with experimental observations.

{\it {Pseudopotentials and Laughlin correlations.}} In this paper we
use Haldane's spherical geometry \cite{HaldanePRL83}, in which there
are no boundary conditions to be imposed and the planar
translational symmetry is replaced by the spherical one. The
perpendicular magnetic field is produced by a monopole with the
strength $2Q$ and the Landau levels $n$ are replaced by angular
momentum shells $\ell$=$Q$+$n$. The Coulomb interaction is described
using the pseudopotentials $V(\mathcal R)$, where $\mathcal R$ ia
the relative angular momentum $\mathcal R= 2\ell -L^{\prime}$,
$L^{\prime}$ being the pair angular momentum. It is well-known that
Laughlin correlations (the avoidance of pair states with small
values of $\mathcal R$) occur only when the pseudopotential
$V_n(\mathcal R)$ describing the interaction energy of a electron
pair with angular momentum $L'=2 \ell -\mathcal R$ in LLn is
``superharmonic", i.e. rises with increasing $L'$ faster than
$L'(L'+1)$ as the avoided value of $L'$ is approached
\cite{QuinnWoysSSC98,WojsQuinnSSC99, QuinnWPhysE00}. In LL0 the
pseudopotential is superharmonic for all values of $\mathcal R$, but
in LL1 it is superharmonic only for $R\geq 3$. Finite well-width
effects can make $V_{1}(R)$ weakly superharmonic at $\mathcal R =1$
\cite{RezayiHaldanePRL00}. However, for well-width of moderate size,
we have found that $V_1(\mathcal R)$ does not change enough to cause
robust Laughlin correlations.  We will ignore finite width effect in
this paper.

When $1/2\geq \nu \geq 1/3$, the lowest energy states in LL0 contain
LCEs (Laughlin correlated electrons) which avoid pairs at $\mathcal
R =1$. A Laughlin-Jain \cite{LaughlinPRL83,JainPRL89} sequence of
integrally filled LCE levels occurs at $\nu=n(2n\pm 1)^{-1}$, when
the composite Fermion (CF) angular momentum
$\ell^{\ast}=|\ell-(N-1)|$ satisfies $2 \ell^{\ast}=N/n-n$. For
$1/3> \nu \geq 1/5$, the LCEs avoid pair states with $\mathcal R=1$
and $3$, giving $\ell^{\ast}=|\ell-2(N-1)|$ and $\nu=n(4n\pm
1)^{-1}$ FQH states.

$V_1(\mathcal R)$ is not superharmonic \cite{WojsPRB01,
WojsQuinnPRB05} at $\mathcal R =1$, so LCEs are not expected for
$1/2\geq \nu \geq 1/3$. Instead, the electrons tend to form pairs
with
\begin{equation}
\label{eq_lp} \ell_{\rm{P}}=2\ell-1~.
\end{equation}
The pairs of electrons can be treated as Bosons or Fermions since in
2D systems a Chern-Simons transformation can change Bosons to
Fermions \cite{QuinnPhysicaE2001}. To avoid violating the exclusion
principle, we can't allow Fermion pairs (FPs) to be too close to one
another. We do this by restricting the angular momentum of two pairs
to values less than or equal to
\cite{QuinnWojsYiPLA03,WWQ2005,WWQPRB06,QuinnQuinnSSC06}
\begin{equation}
\label{eq_lFP} 2\ell_{\rm{FP}}=2\ell_{\rm{P}}-3(N_{\rm{P}}-1)~,
\end{equation}
where the number of pairs is:
\begin{equation}
\label{eq_N_P} N_{\rm{P}}=\frac{N}{2}~.
\end{equation}
%If the pairs were treated as Bosons, we would take $2\ell_{\rm{BP}}=2\ell_{\rm{P}}-4(N_{\rm{P}}-1)$.
From this it is apparent that the FP filling factor satisfies the
relation $\nu_{\rm{FP}}^{-1}=4\nu_{1}^{-1}-3$. The factor of 4 is a
reflection of $N_{\rm{P}}$ being half of $N$ and the LL degeneracy
$g_{\rm{P}}$ of the pairs being twice $g$ for electrons.
Correlations are introduced through a standard CF transformation
applied to FPs:
\begin{equation}
2\ell_{\rm{FP}}^{\ast}=2\ell_{\rm{FP}}-2p(N_{\rm{P}}-1).
\label{eq_lFP_star}
\end{equation}

\begin{figure}
\centerline{\includegraphics[width=0.52
\linewidth]{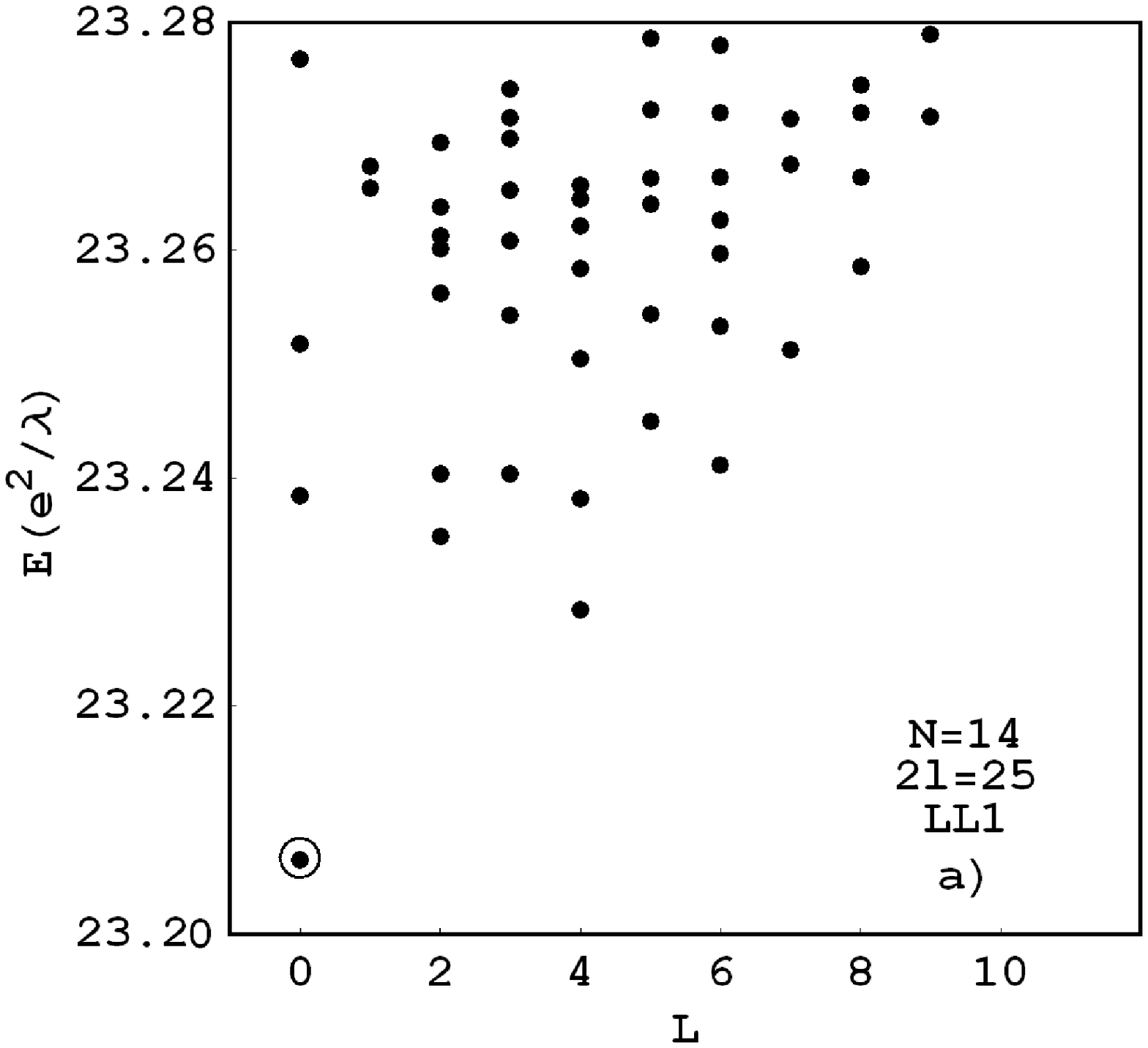}
\includegraphics[width=0.48
\linewidth]{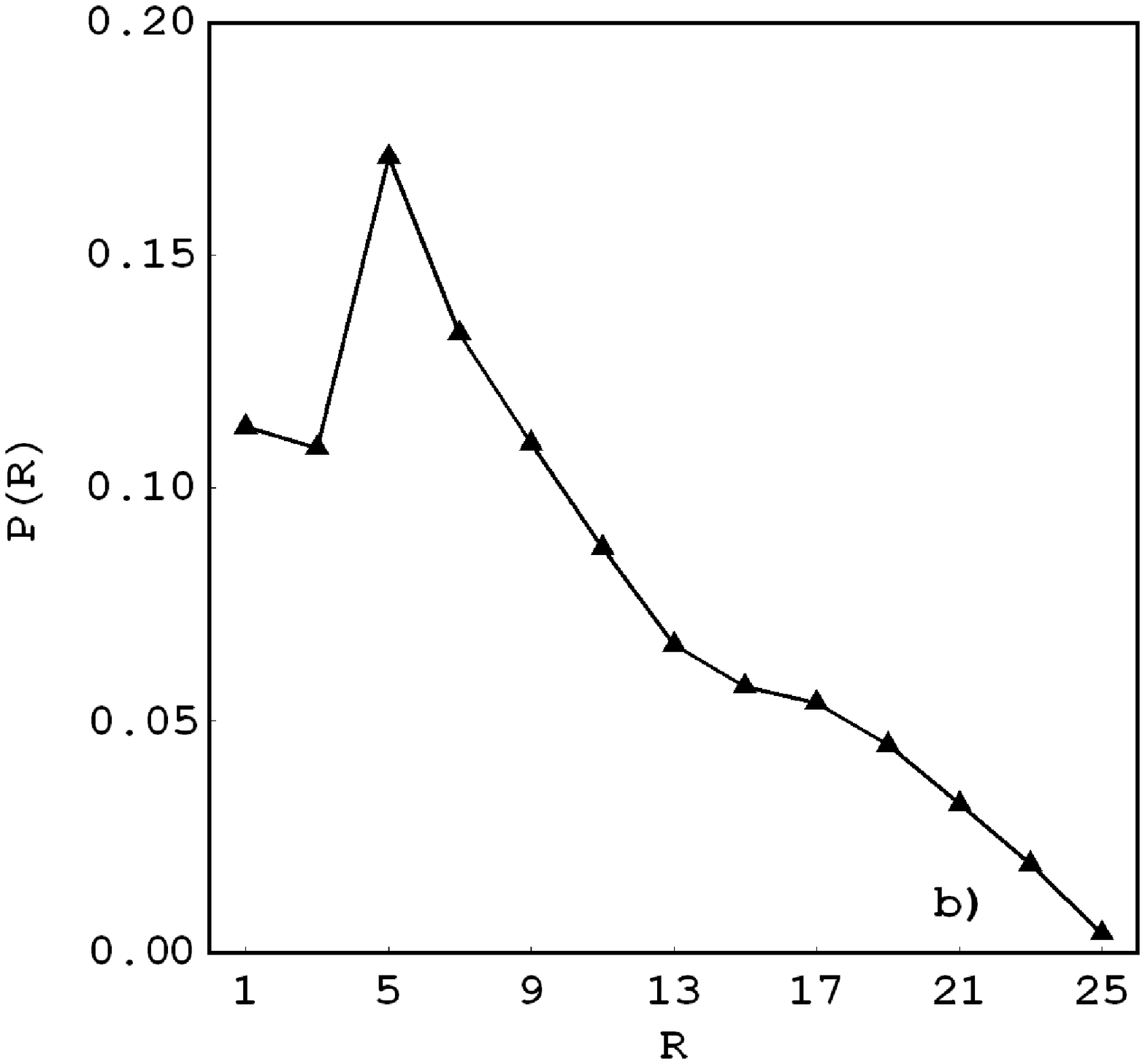}} \caption{\label{Fig14_25}a) The
low energy spectrum of $N=14$ electrons in LL1 at $2\ell=25$. The
$L=0$ IQL state, marked by a circle, is separated from higher state
by a gap. b) The probability $P(\mathcal R)$ of pair for $L=0$
ground state with relative angular momentum $\mathcal R$ as a
function of $\mathcal R$. $P(\mathcal R)$ is not a minimum at
$\mathcal R=1$, and not a maximum at $\mathcal R=3$ as in a Laughlin
state.}
\end{figure}

For $(2\ell,N)=(25,14)$ and $2p=4$ this gives $\ell_{\rm{P}}=24$,
$N_{\rm{P}}=7$, $2\ell_{\rm{FP}}=30$ and $2\ell_{\rm{FP}}^{\ast}=6$.
The $N_{\rm{P}}=7$ FPs fill the $\ell_{\rm{FP}}^{\ast}=3$ shell
giving an $L=0$ IQL ground state with FP filling factor
$\nu_{\rm{FP}}=(2p+1)^{-1}=1/5$ and $\nu_1=1/2$. The numerical
spectrum is shown in Figure \ref{Fig14_25} a); it has a well defined
gap separating the ground state from the excited states. $P(\mathcal
R)$ vs. $\mathcal R$ is obtained from the eigenfunctions and
contains the same information for a spherical surface as the pair
distribution function on a plane. The maximum at $\mathcal R=1$ and
minimum at $\mathcal R=3$ is incontrovertible evidence that the
electrons are not Laughlin correlated. Our $L=0$ ground state of
Laughlin correlated pairs is definitely  different form the
Moore-Read Pfaffian state \cite{MooreReadNuclPhys91}, the exact
eigenstate of a special three particle repulsive interaction
\cite{GreiterWenWilczekPRL91,WojsQuinnPRB05}. The square of the
overlap of these two ground state wavefunctions is only 0.48; the
overlap of the excited states for the two models is much smaller.

%When $\nu_1=1/2$, $\nu_{\rm{FP}}=1/5$. This decreased
%$\nu_{\rm{FP}}$ allows the FPs to be Laughlin correlated, with
%$2\ell_{\rm{FP}}=5(N_{\rm{P}}-1)$ giving $2\ell=2N-3$, where $\ell$
%and $N$ refer to the angular momentum shell and number of electrons
%in LL1. Figure \ref{Fig14_25} a) gives the energy spectrum for
%$N=14$ electrons with $2\ell=25$ in LL1, using the Coulomb
%pseudopotential appropriate for a quantum well with zero width. The
%$L=0$ ground state is separated from the excited states by a
%well-defined gap. This state is definitely different form the
%Moore-Read \cite{MooreReadNuclPhys91} Pfaffian state as the overlap
%between with our state is only 0.48.  In frame b), we plot
%$P(\mathcal R)$ vs. $\mathcal R$ for the ground state. $P(\mathcal
%R)$ is obtained from the eigenfunction and contains the same
%information for the spherical surfaces as the pair distribution
%function on a plane. The fact that $P(\mathcal R)$ is not a minimum
%at $\mathcal R=1$, and that it is a maximum at $\mathcal R=5$, not
%$\mathcal R=3$, is incontrovertible evidence that the electrons are
%not Laughlin correlated.
%$P(\mathcal R)$ for the LCE state of LL0 at $2\ell=33$ and $N=12$ is also shown
% in frame b) to emphasize the difference.

\begin{figure}
\centerline{\includegraphics[width=0.5 \linewidth]{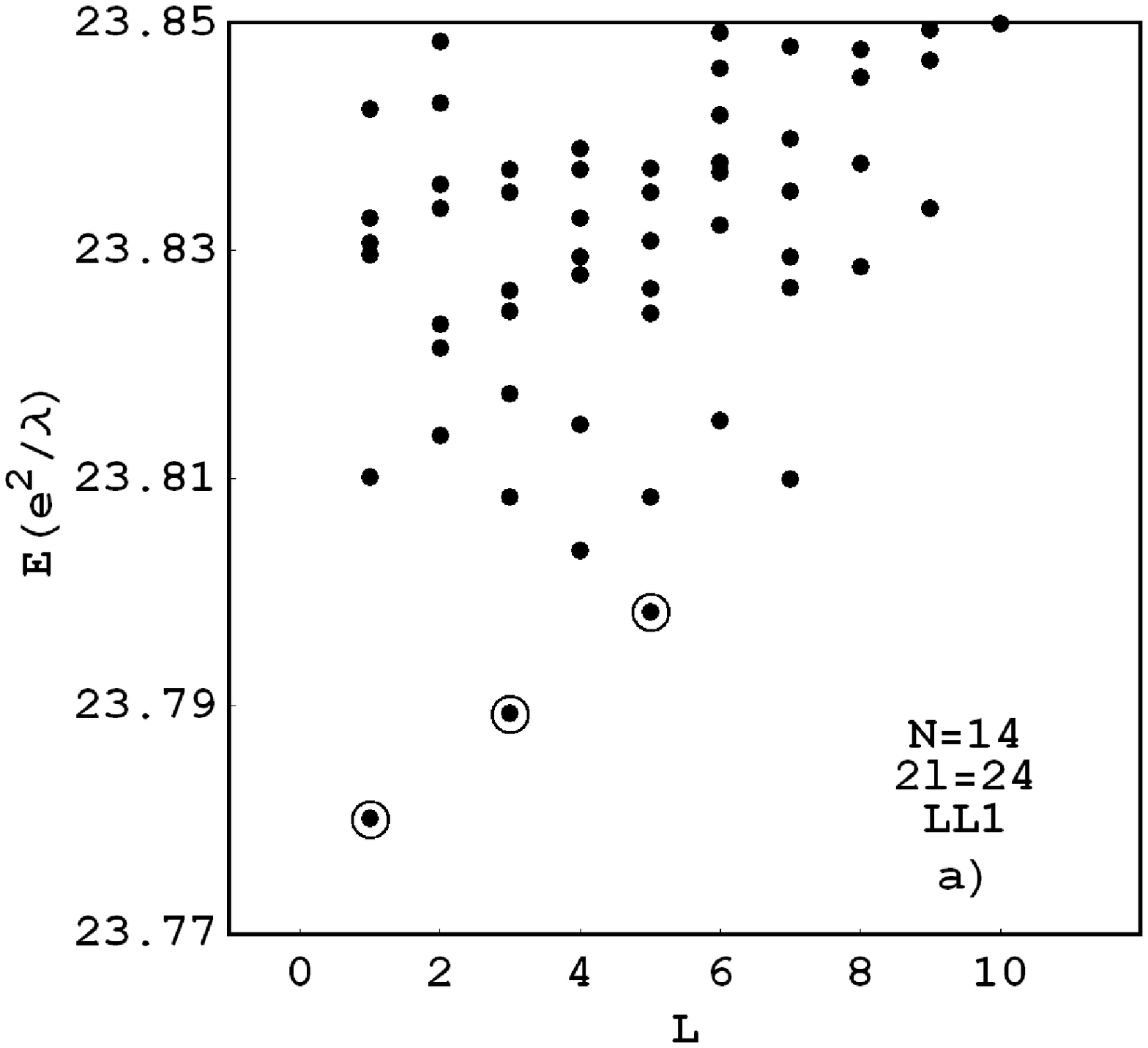}
\includegraphics[width=0.5 \linewidth]{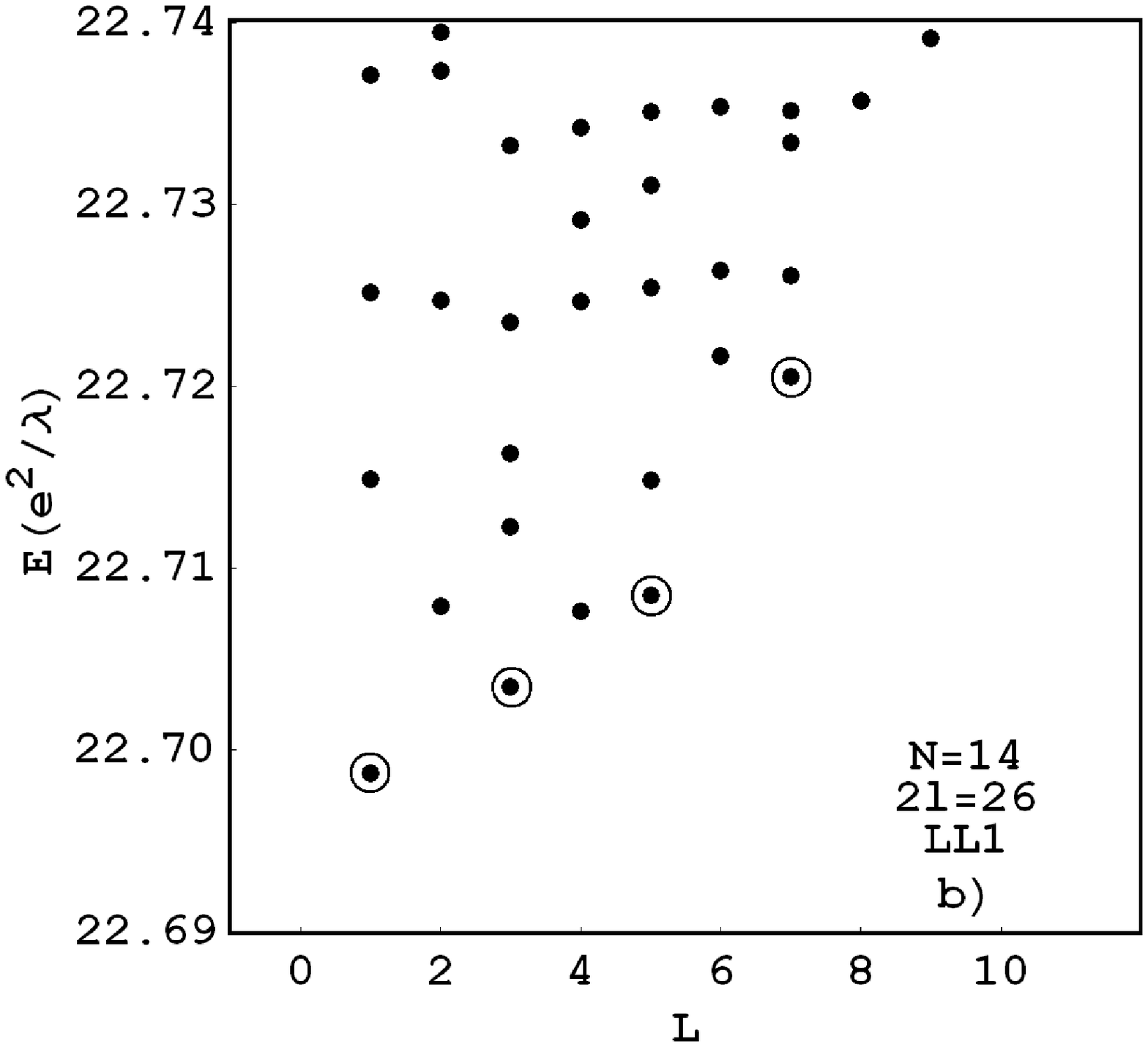}}
\caption{\label{Fig14_24_26}a) The low energy spectrum of $N=14$
electrons in LL1 at $2\ell=24$.  b)  The low energy spectrum of
$N=14$ electrons in LL1 at $2\ell=26$. Changing $2\ell=2N-3$ by one
unit results in a pair of QP or QH excitations, marked in figure
with circles.}
\end{figure}

In Figure \ref{Fig14_24_26} we present the two energy spectra for
$N=14$ and $2\ell=24$ (a) and $26$ (b). Increasing or decreasing
$2\ell$ (from the values that give an IQL ground state) by one unit
clearly produces a pair of elementary excitations and a band of low
lying states. We can understand these spectra using our set of
equations (\ref{eq_lp}-\ref{eq_lFP_star}) for paired states with
$(2\ell,N)=(24,14)$ and (26,14). The former case contains two FP
quasiparticles each with $\ell_{\rm{FPQP}}=3$, and the latter two FP
quasiholes with $\ell_{\rm{FPQH}}=4$, giving the lowest energy bands
marked by circled dots in Fig. \ref{Fig14_24_26}. The occurrence of
FQH states at $2\ell=2N-3$ and at the electron-hole conjugate
$2\ell=2N+1$ (obtained by replacing $N$ with $2 \ell +1-N$ ) agrees
with numerical calculations. The gap appears to approach a finite
limit for large $N$ when plotted as a function of $N^{-1}$, but our
calculations are limited $8\leq N \leq 16$. A meaningful comparison
with experimental data would require taking into account the finite
well-width, the effect of disorder and especially Landau level
mixing as suggested by very recent experimental data
\cite{LL1_Exp_5_2}. The low lying states for values of $2\ell$ close
to $2N-3$ seem to be well described in terms of Laughlin correlated
pairs of electrons with the pairs treated as Fermions.

\begin{figure}
\begin{center}
\centerline{
\includegraphics[width=0.5 \linewidth]{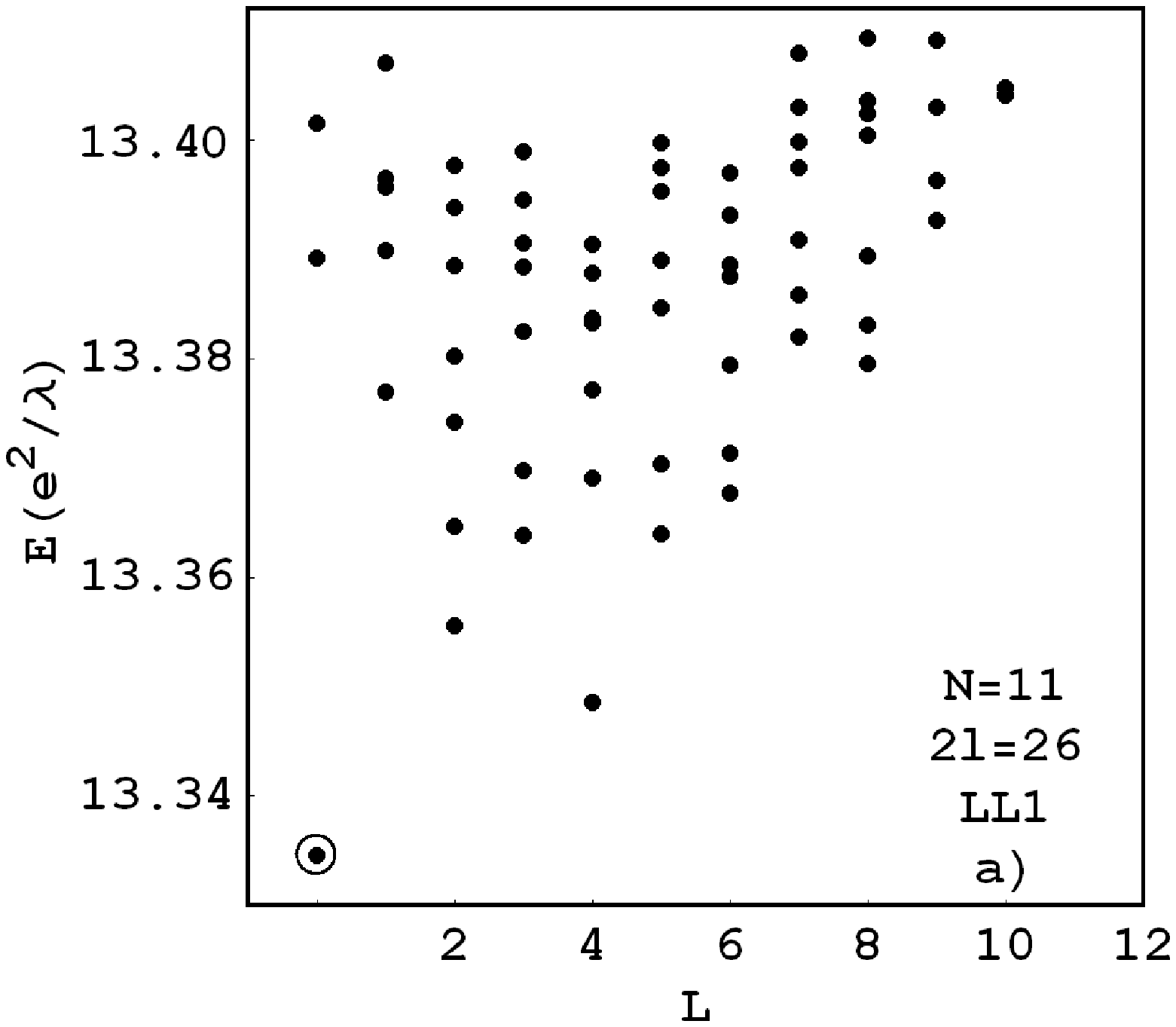},\includegraphics[width=0.5
\linewidth]{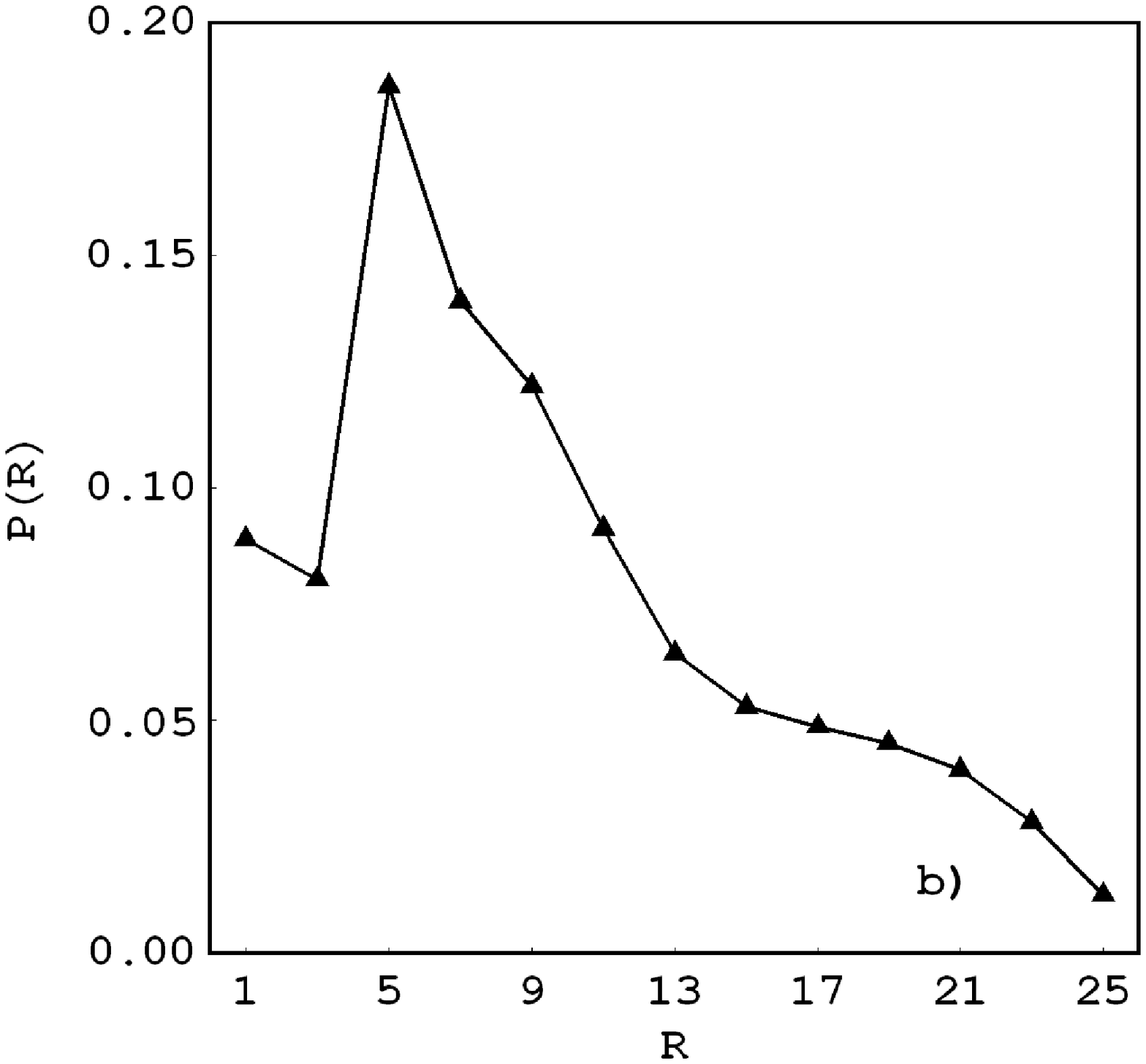}} \caption{\label{Fig12_29} a)
Spectrum of $N=11$ electrons at $2 \ell =26$ in LL1. $L=0$ ground
state is marked by a circle. b)$P(\mathcal R)$ vs. $\mathcal R$ for
the ground state. For $\mathcal R=1$, $P(\mathcal R)$ is not a
minimum, but for $\mathcal R=3$ it is. This is exactly opposite to
the behavior in LCE state.}
\end{center}
\end{figure}

For $\nu_1=1/3$ in LL1, we performed numerical calculations for many
different values of $2\ell$ for each value of $N$ \cite{WojsPRB01,
WojsQuinnPRB05}. $L=0$ ground states were found for $6\leq N \leq
12$ at $2\ell=3N-7$, quite a different value than for the $L=0$
incompressible state in LL0, which occur at $2\ell=3N-3$. Although
pairs are expected to form, complete pairing for even values of $N$
would give an incompressible ground state at $2\ell=3N-5$, not
$3N-7$. We do not fully understand the correlations at $\nu_1=1/3$;
larger cluster than pairs could form or we could have a plasma with
clusters of several different sizes. Such systems might be treated
by a generalized composite Fermion picture
\cite{WojsSzYiQuinnPRB99}, but we do not consider a plasma with two
or more different cluster sizes in the present paper. In Figure
\ref{Fig12_29} we show an example of a spectrum, $N=11$ and
$2\ell=26$, and a plot of $P(\mathcal R)$ vs. $\mathcal R$ for the
$L=0$ ground state. It is clear that the electrons are not Laughlin
correlated since $P(R)$ is a minimum for $\mathcal R=3$, not
$\mathcal R=1$. Unfortunately, the gap size is not a smooth function
of $N^{-1}$, so our limited range of $N$ values does not allow
extrapolation to the macroscopic limit with any degree of
confidence. We are certain however that pairs or larger clusters
will form instead of LCEs.

\begin{figure}
\begin{center}
\centerline{
\includegraphics[width=0.5 \linewidth]{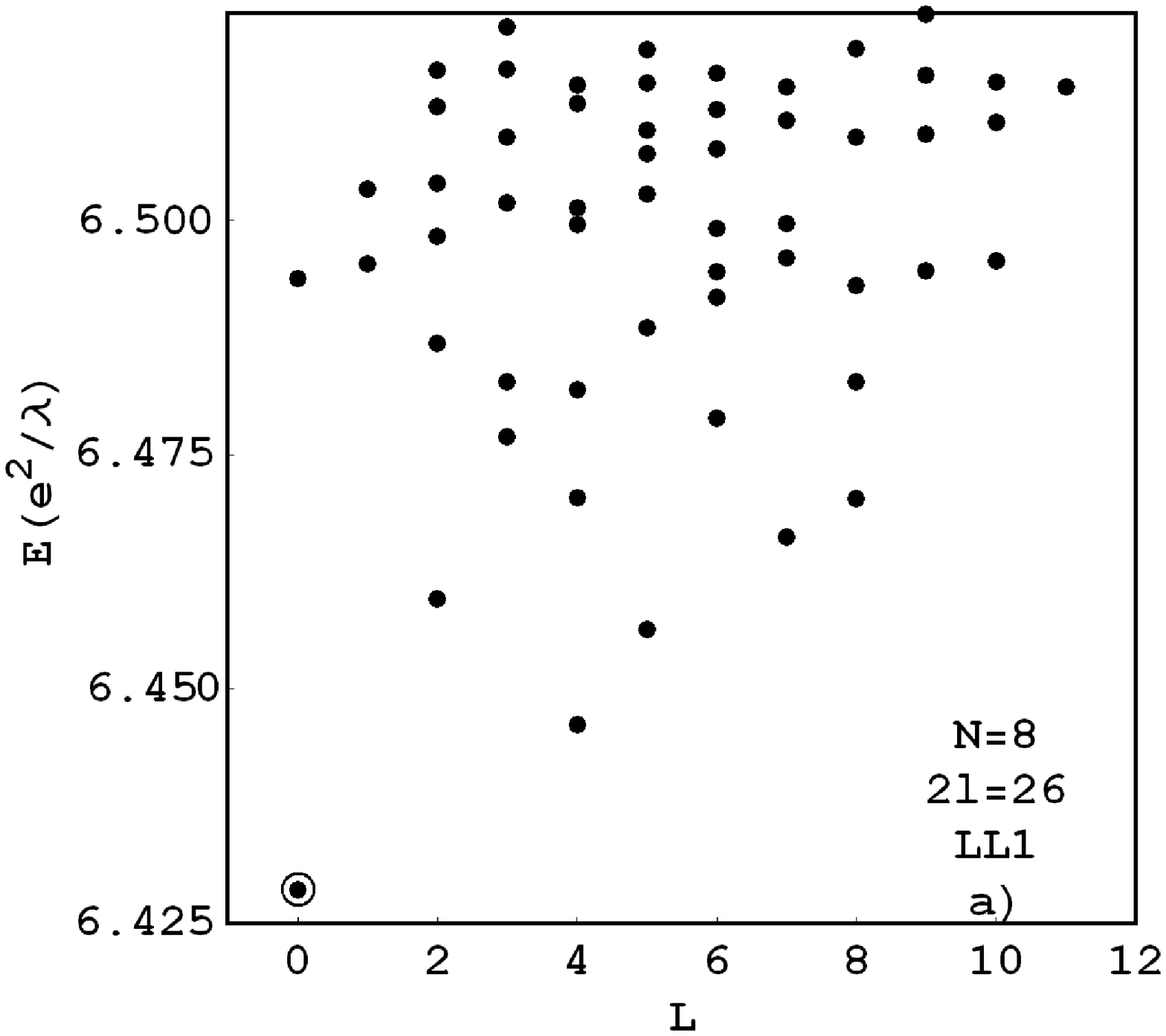}
\includegraphics[width=0.5 \linewidth]{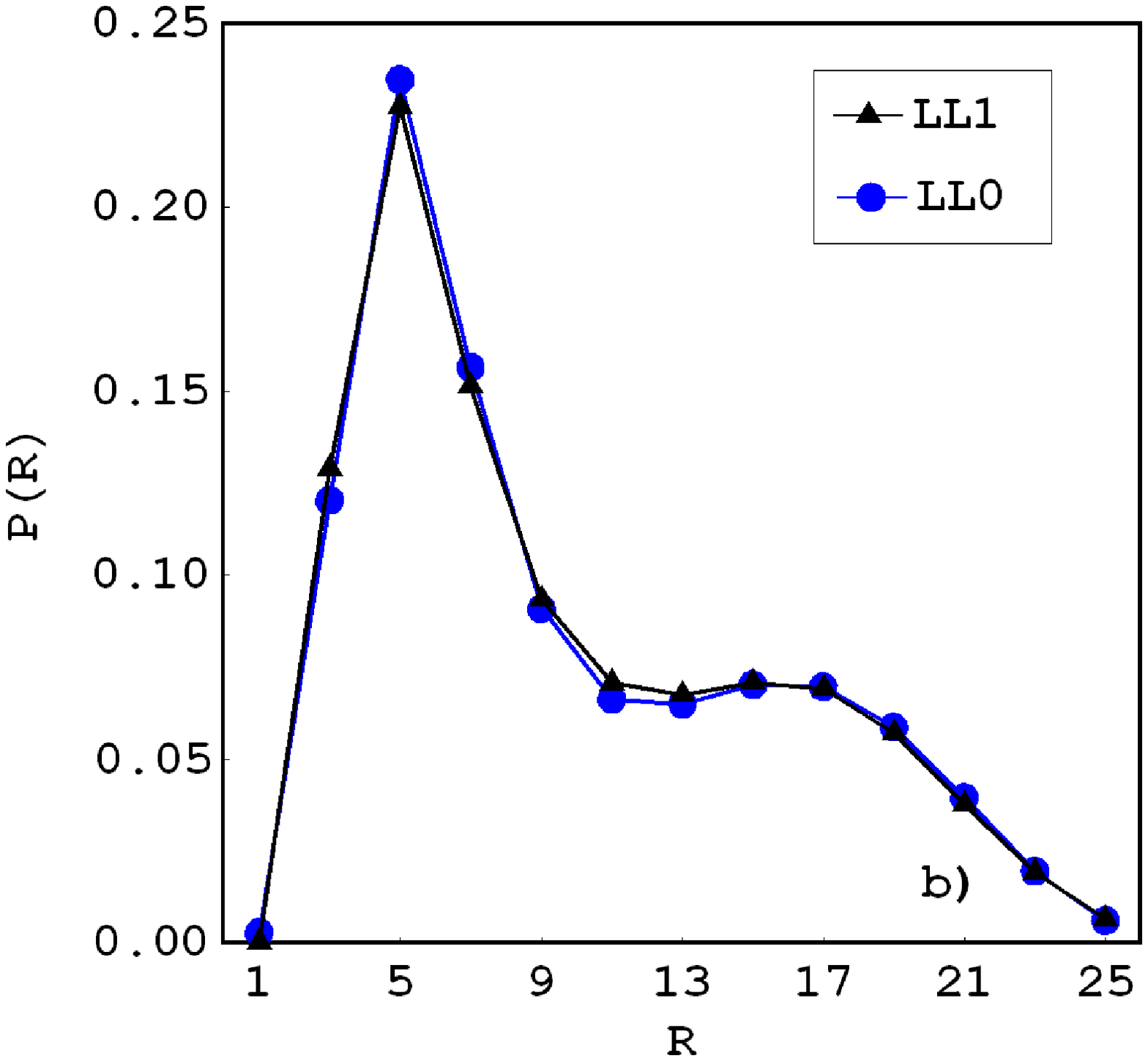}}
\caption{\label{Fig8_26} (color online) a) Energy spectrum for $N=8$
electrons at $2 \ell =26$ iin LL1. b) $P(\mathcal R)$ vs. $\mathcal
R$ for the $L=0$ ground state. Dashed line is for $P(\mathcal R)$
vs. $\mathcal R$ in the LL0.}
\end{center}
\end{figure}

For $1/3 > \nu_1 \geq 1/5$, Laughlin correlations with $\rm{CF}^4$
(composite Fermions with 4 attached flux quanta) are expected, just
as in LL0. The spectrum for $N=8$ and $2\ell=\frac{7}{2}N-2=26$, and
$P(\mathcal R)$ vs. $\mathcal R$ for the ground state are shown in
Fig.\ref{Fig8_26}. The low values of $P(\mathcal R)$ at $\mathcal
R=1$ and $3$ agree with the $\rm{CF}^4$ picture giving a Jain state
at $\nu_1=n(4n-1)^{-1}$ with $n=2$. The $L=0$ ground state is marked
with a circle. The LCEs have angular momentum
$\ell^{\ast}=\ell-2(N-1)=1$. The lowest LCE level can hold
$2\ell^{\ast}+1=3$ CFEs; the remaining 5 fill the QE level with
$\ell_{\rm{QE}}=2$ giving the daughter state at $\nu_1=2/7$. The
correlations are similar to those in the $\nu=2/7$ state in LL0.

\begin{figure}
\begin{center}
\centerline{
\includegraphics[width=0.5 \linewidth]{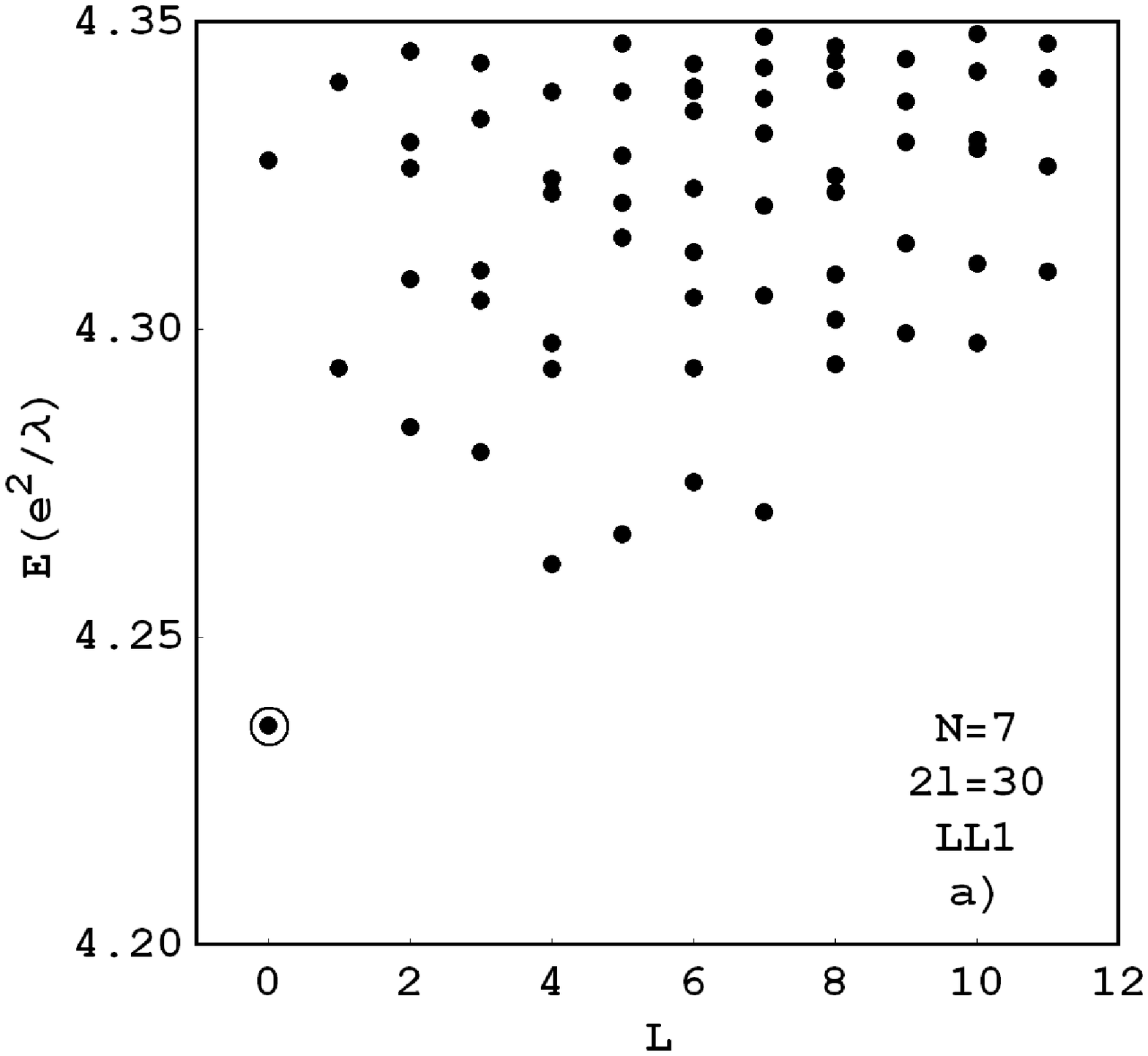}
\includegraphics[width=0.5 \linewidth]{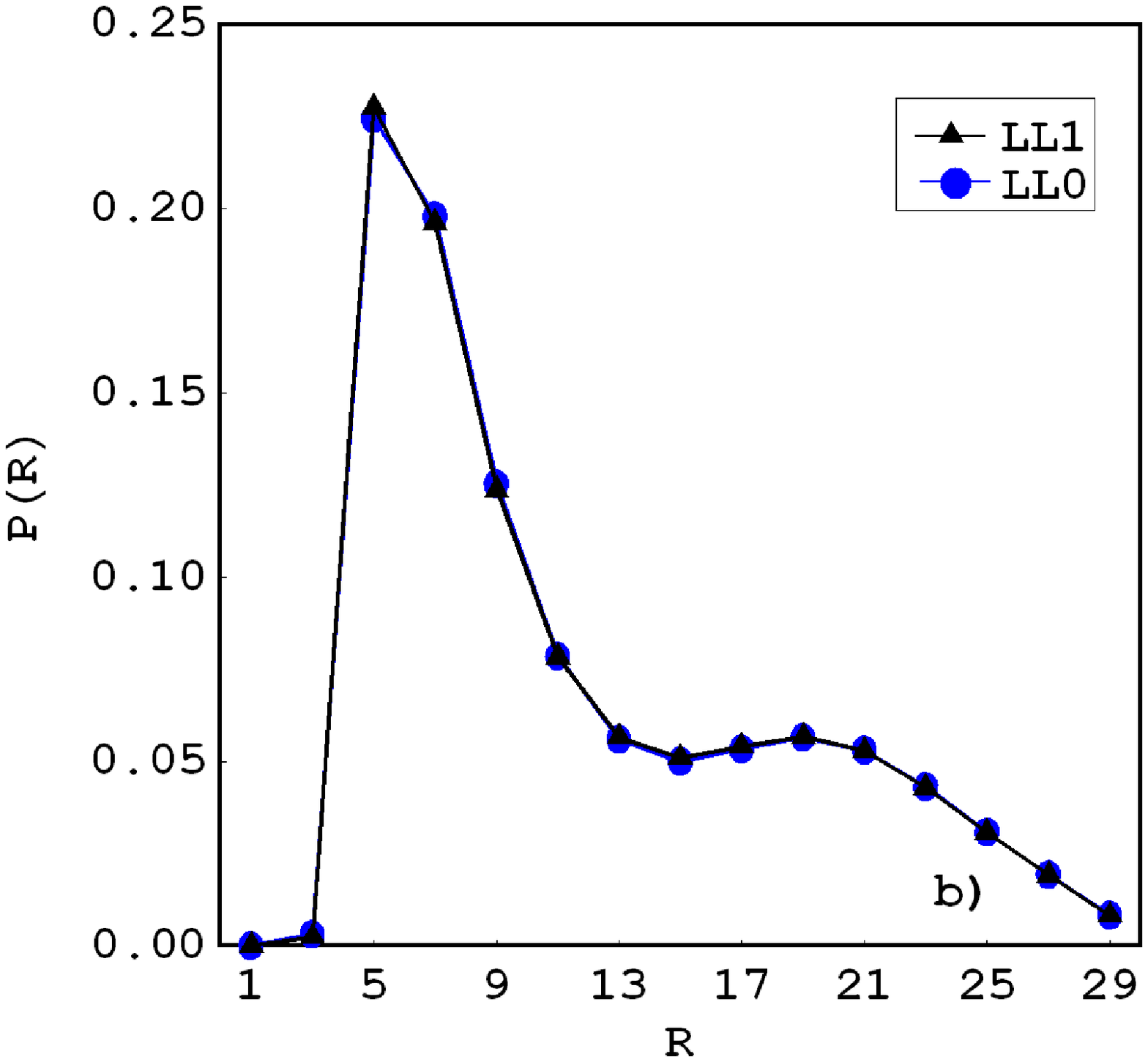}}
\caption{\label{EnSpect1_5} (color online) a) Spectrum of a system
with 7 electrons at $2 \ell =30$ in the first excited Landau level
b) $P(\mathcal R)$ vs. $\mathcal R$ for the $L=0$ ground
state.Dashed line for $P(\mathcal R)$ vs. $\mathcal R$ in the LL0,
is essentially indistinguishable from solid line.}
\end{center}
\end{figure}

For $\nu_1=1/5$ the ground state is also an incompressible $L=0$
state which avoids pair states with $\mathcal R=1$ and $3$, similar
to the $1/5$ state in LL0. The energy spectrum and pair probability
for the incompressible ground state at $L=0$ are presented in Fig.
\ref{EnSpect1_5}. The pair probability $P(\mathcal R)$ for $\nu=1/5$
and $\nu=2/7$ states in LL0 are also shown in Figures \ref{Fig8_26}
b) and \ref{EnSpect1_5} b) to emphasize the similarity of LCE states
in LL0 and LL1.

\begin{figure}
\begin{center}
\centerline{
\includegraphics[width=0.5 \linewidth]{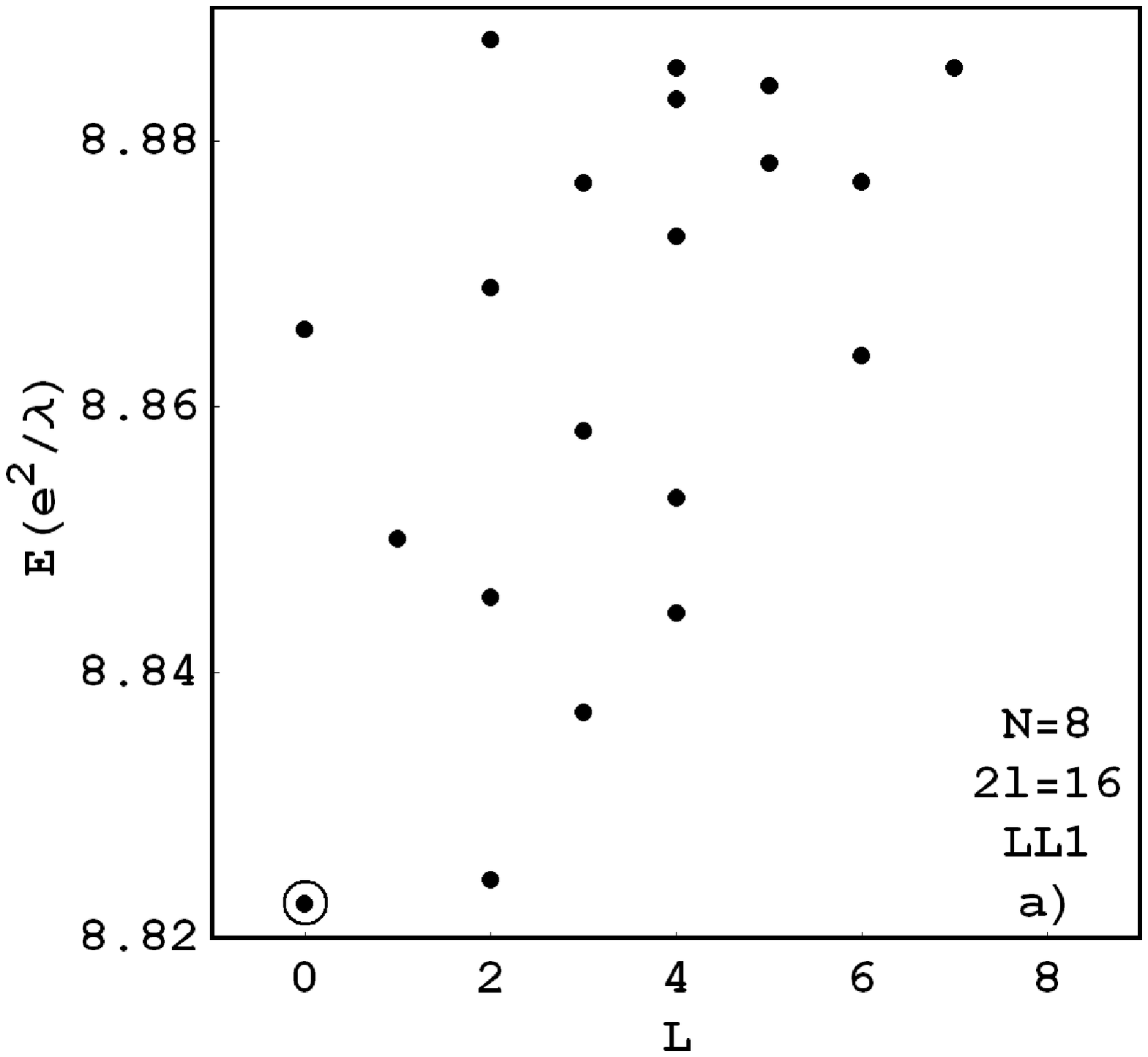}
\includegraphics[width=0.5 \linewidth]{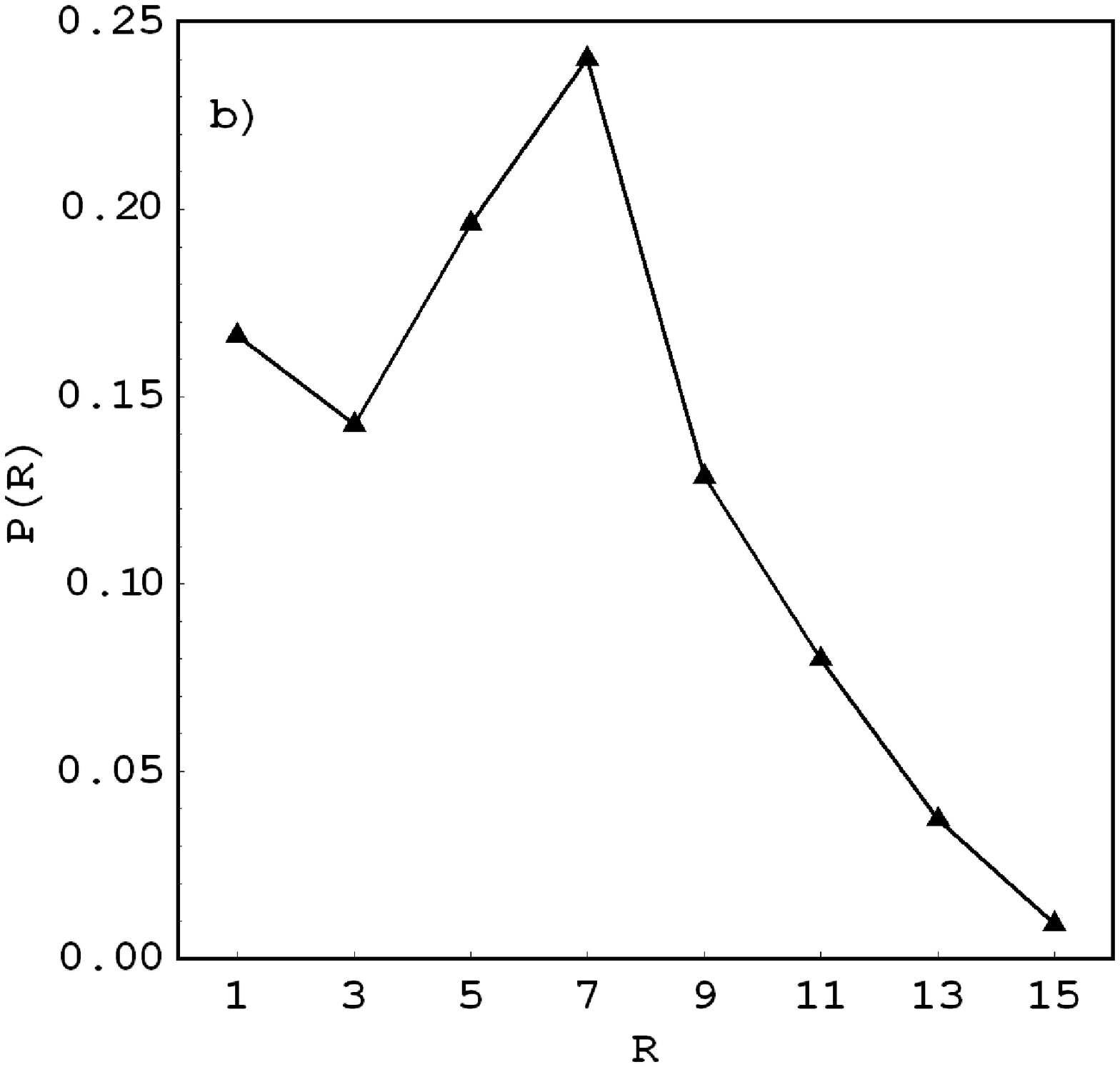}}
\caption{\label{Fig2_5} a) Spectrum of a system with 8 electrons at
$2 \ell =16$ in the first excited Landau level. The enrgy gap is
very small b) $P(\mathcal R)$ vs. $\mathcal R$ for the $L=0$ ground
state.}
\end{center}
\end{figure}

The FQH state at $\nu_1=2/5$ has been experimentally identified by
Xia et al. \cite{Pan_LL1} and reconfirmed recently \cite{LL1_Exp},
but it still lacks a clear theoretical explanation. Because the
pseudopotential $V_1(\mathcal R)$ is not superharmonic at $\mathcal
R=1$, the electrons are not Laughlin correlated. The series of
$2\ell=\frac{5}{2}N-4$ produces $L=0$ incompressible ground state
for even $N \leq 12$. In Figure \ref{Fig2_5} we present the energy
spectrum (frame a) and the pair probability (frame b) of a system
with 8 electrons and $2\ell=16$. The pair probability clearly looks
different from what one expects from a Laughlin correlated state.

\begin{figure}
\begin{center}
\centerline{
\includegraphics[width=0.5 \linewidth]{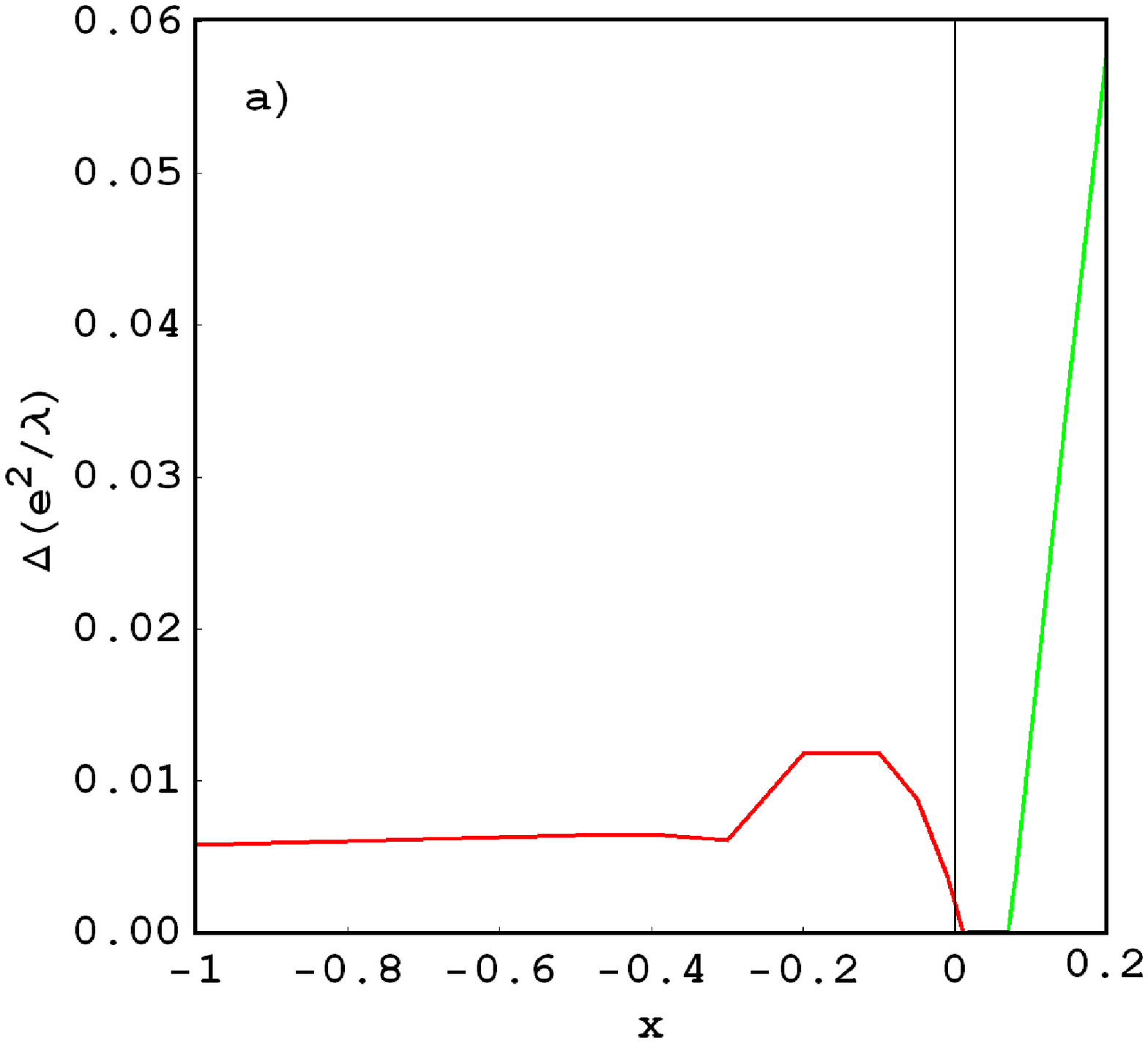}
\includegraphics[width=0.5 \linewidth]{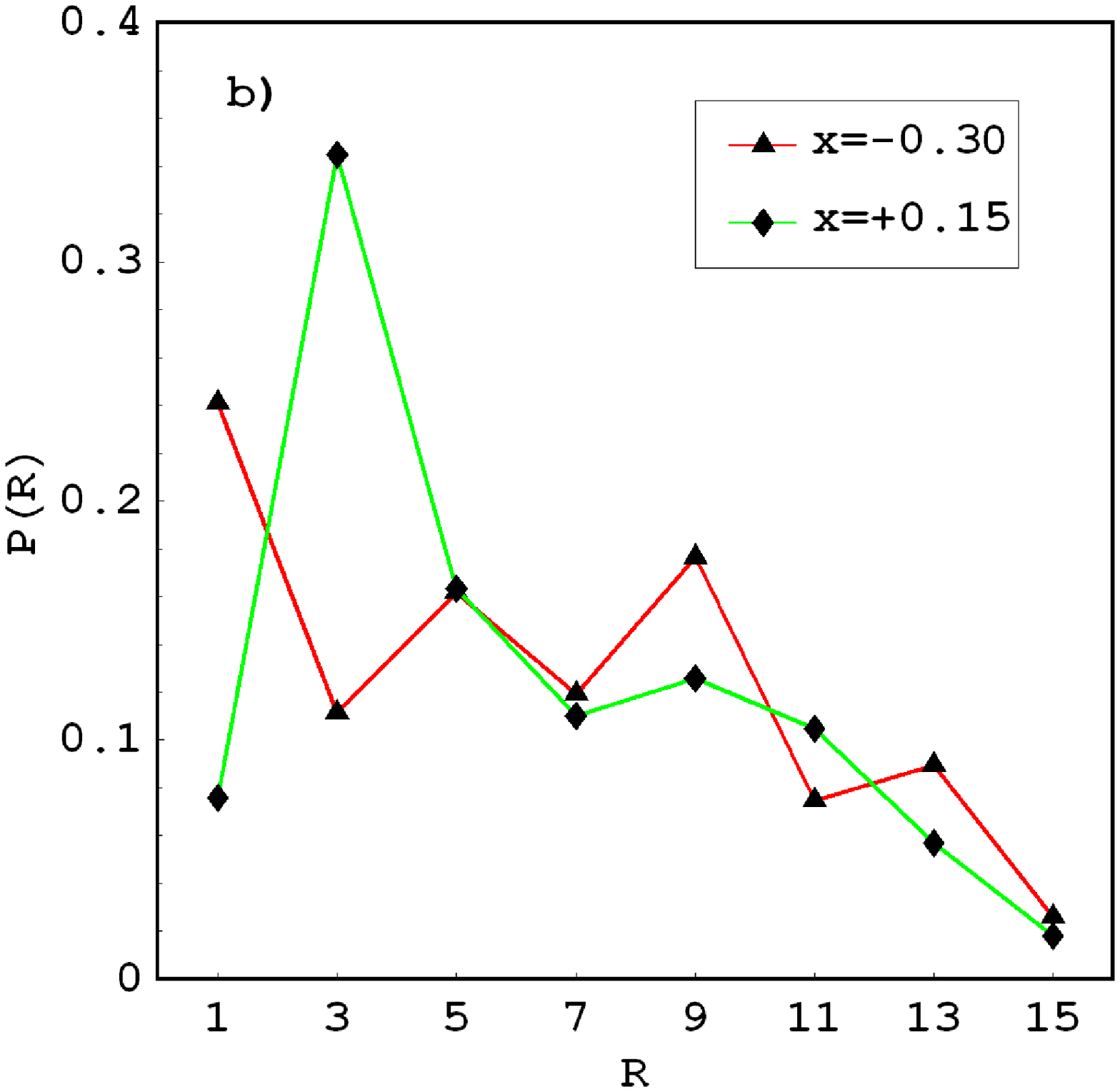}}
\caption{\label{Fig2_5_gap_pr} (color online) a) Energy gap vs.
$\delta V_1/V_1=x$. Remainder of pseudopotential $V_1(\mathcal R)$
(for $\mathcal R=3,5,\ldots$) is unchanged. b) Sketch of pair
probability for $x=-0.3$ (red) and $x=0.15$(green)}
\end{center}
\end{figure}

It is interesting to investigate the $\nu_1=2/5$ state in LL1, when
only $V_1(\mathcal R=1)$ (or simply $V_1(1)$) is changed from its
Coulomb value in a zero width quantum well by an amount $\delta
V_1=xV_1(1)$, while all other values of $V_1(\mathcal R)$ are
unchanged. In Fig. \ref{Fig2_5_gap_pr} a) we show the energy gap
$\Delta$ (when it is positive) between the $L=0$ state and the
closest excited state as a function of $x$ for an eight electron
system \cite{note_state_2_5} at $2\ell=16$. For $x \lesssim -0.3$ a
very small gap ($\simeq 0.0065 e^2/\lambda$) occurs. As $x$
increases this gap first increases slightly and then begin to
decrease at $x \simeq -0.1$. It disappears at $x \simeq 0.01$ but
reappears for $x \gtrsim 0.08$, increasing almost linearly with $x$
for larger $x$. In frame b) we present $P(\mathcal R)$, the
probability of having pairs with relative angular momentum $\mathcal
R$ for the ground state for two values of $x$, $x=-0.3$ and
$x=0.15$. The latter is clearly superharmonic (as in LL0), giving a
robust Laughlin-Jain state (LJS) at $\nu_1=2/5$. The former is
subharmonic giving a small gap related to the formation of a
Laughlin correlated state of pairs of electrons (LCP). Both of these
states occur for $2\ell=\frac{5}{2}N-4$. For zero well-width LL1
corresponds to $\delta V_1$ equal to zero. Finite well width effects
seem too small to change its non-Laughlin correlated behavior. Thus
we expect a robust $\nu=2/5$ LJS in LL0, but at most a very small
gap LCP state in LL1.

{\it Summary.} Because $V^{(1)}(\mathcal R)$ in LL1 is not
superharmonic at $\mathcal R=1$, but is at $\mathcal R=3$, the
ground state can not be Laughlin correlated for the filling factors
$\nu_1=\nu -2$ in the range $1/2 \geq \nu_1 \geq 1/3$,  but it can
be for $1/3> \nu_1 \geq 1/5$. The CF picture with 2 attached flux
quanta ($\rm{CF}^2$) cannot be applied. The FQH states at
$\nu_1=2/3$, $1/2$, $1/3$ must contain pairs with
$\ell_{\rm{P}}=2\ell-1$ (or larger clusters). In general, pairing or
formation of larger clusters give weaker FQH states than Laughlin
correlations.

For $2\ell=2N-3$ or $2N+1$ in LL1 and $N$ even,
$N_{\rm{P}}=\frac{1}{2} N$ pairs can form a Laughlin correlated
state of pairs (LCP state). Numerical calculations confirm $L=0$
ground states at $\nu=2+\nu_1=5/2$ for $2\ell=2N-3$ or $2N+1$ with
robust gaps. For $\nu_1=1/3$ (and its e-h conjugate at $\nu_1=2/3$)
the correlations aren't understood. FQH states are formed for $6
\leq N \leq 12$ at $2\ell=3N-7$. LCEs in LL0 have $2\ell=3N-3$, and
Laughlin correlated pairs in LL1 would occur at $2\ell=3N-5$.
Through LC triplets would occur for $N$ a multiple of $3$ at
$2\ell=3N-7$, we don't have enough numerical data to be certain of
correlations at $\nu_1=1/3$. Mixed plasmas of different size
clusters may be necessary to understand the correlations. For $1/3 >
\nu_1 \geq 1/5$, the LL1 electrons can be Laughlin correlated. LCE
states have larger gaps than the LC states of pairs (or larger
clusters). This leads to Laughlin-Jain states at $\nu_1=\frac{n}{4n
\pm 1}$ explaining why $\nu_1=1/5$ , $2/7$ states and their e-h
conjugate are prominent but $\nu_1=2/5$ isn't.

{\it Acknowledgements.} The authors wish to acknowledge the support
of the Basic Energy Sciences Program of the DOE, and to thank to S.
Das Sarma and A. W\'ojs for useful discussions.

%\bibliography{LL1PRB}

\begin{thebibliography}{29}
\bibitem[{not()}]{note_state_2_5}
\bibinfo{note}{The case of $N=10$, $2\ell=21$ has also been evaluated with
  similar results. It is not shown because it is an 'aliased' one and can be
  attributed to both $\nu_1=1/2$ ($2\ell =2N+1$) and $\nu_1=2/5$ ($2\ell
  =\frac{5}{2}N-4$)}.
\end{thebibliography}

\end{document}